\documentclass[aps,prl,twocolumn,superscriptaddress,letterpaper]{revtex4-2}

\usepackage{graphicx}
\usepackage{bm}
\usepackage{amssymb}
\usepackage{color}
\usepackage{amsmath}
\usepackage[colorlinks=true,linkcolor=blue,anchorcolor=black,citecolor=blue,filecolor=black,menucolor=black,runcolor=black,urlcolor=blue]{hyperref}

\def \Z {\mathbb{Z}}
\def \L {\mathsf{L}}

\begin{document}

\title{Projectively enriched symmetry and topology in acoustic crystals}

\author{Haoran Xue}
\thanks{These authors contributed equally.}
\affiliation{Division of Physics and Applied Physics, School of Physical and Mathematical Sciences, Nanyang Technological University,
Singapore 637371, Singapore}

\author{Zihao Wang}
\thanks{These authors contributed equally.}
\affiliation{Division of Physics and Applied Physics, School of Physical and Mathematical Sciences, Nanyang Technological University,
Singapore 637371, Singapore}

\author{Yue-Xin Huang}
\affiliation{Research Laboratory for Quantum Materials, Singapore University of Technology and Design, Singapore 487372, Singapore}

\author{Zheyu Cheng}
\affiliation{Division of Physics and Applied Physics, School of Physical and Mathematical Sciences, Nanyang Technological University,
Singapore 637371, Singapore}

\author{Letian Yu}
\affiliation{Division of Physics and Applied Physics, School of Physical and Mathematical Sciences, Nanyang Technological University,
Singapore 637371, Singapore}

\author{Y. X. Foo}
\affiliation{Division of Physics and Applied Physics, School of Physical and Mathematical Sciences, Nanyang Technological University,
Singapore 637371, Singapore}

\author{Y. X. Zhao}
\email[]{zhaoyx@nju.edu.cn}
\affiliation{National Laboratory of Solid State Microstructures and Department of Physics, Nanjing University, Nanjing 210093, China}
\affiliation{Collaborative Innovation Center of Advanced Microstructures, Nanjing University, Nanjing 210093, China}

\author{Shengyuan A. Yang}
\email[]{shengyuan\_yang@sutd.edu.sg}
\address{Research Laboratory for Quantum Materials, Singapore University of Technology and Design, Singapore 487372, Singapore}

\author{Baile Zhang}
\email{blzhang@ntu.edu.sg}
\affiliation{Division of Physics and Applied Physics, School of Physical and Mathematical Sciences, Nanyang Technological University,
Singapore 637371, Singapore}
\affiliation{Centre for Disruptive Photonic Technologies, Nanyang Technological University, Singapore 637371, Singapore}

\maketitle

\textbf{Symmetry plays a key role in modern physics, as manifested in the revolutionary topological classification of matter in the past decade \cite{chiu2016}. So far, we seem to have a complete theory of topological phases from internal symmetries as well as crystallographic symmetry groups \cite{zhang2019, vergniory2019, tang2019}. However, an intrinsic element, i.e., the gauge symmetry in physical systems, has been overlooked in the current framework. Here, we show that the algebraic structure of crystal symmetries can be projectively enriched due to the gauge symmetry, which subsequently gives rise to new topological physics never witnessed under ordinary symmetries. We demonstrate the idea by theoretical analysis, numerical simulation, and experimental realization of a topological acoustic lattice with projective translation symmetries under a $\Z_2$ gauge field, which exhibits unique features of rich topologies, including a single Dirac point, M\"{o}bius topological insulator and graphene-like semimetal phases on a rectangular lattice. Our work reveals the impact when gauge and crystal symmetries meet together with topology, and opens the door to a vast unexplored land of topological states by projective symmetries.}

The concepts of symmetry and topology have permeated into almost all branches of physics \cite{hasan2010, qi2011, ozawa2019, ma2019, huber2016, cooper2019}. In topological matters, the ``topology" refers to the nontrivial global structures in the phase winding of wave functions in momentum space. Symmetries constrain the possible topological structures \cite{chiu2016}, and it is clear that their significance lies in how the symmetries act on the wave functions, or in mathematical terms, how they are represented in the wave function space.

There are two important points that have not been appreciated in the study of topological states of matter. First, under a gauge field, crystal symmetries will be projectively represented, and this impacts the algebra of symmetry operations \cite{zhao2021, zhao2020, shao2021}. This can be understood from the analogy with Aharonov-Bohm effect \cite{aharonov1959}: the phase of wave function for different path is modified by the gauge field, which in turn revises the representation of spatial symmetries. Second, the gauge fields can actually be intrinsic and ubiquitous for real physical systems, not necessarily the applied magnetic field as in the Aharonov-Bohm effect. Particularly, the $\Z_2$ gauge field is intrinsic to artificial periodic systems with time reversal symmetry $T$, i.e., the hopping amplitudes are real numbers that can take either positive or negative signs, and moreover, it can be precisely engineered with current technology \cite{keil2016, serra2018, peterson2018, xue2020, ni2020, qi2020}.

\begin{figure*}
  \centering
  \includegraphics[width=\textwidth]{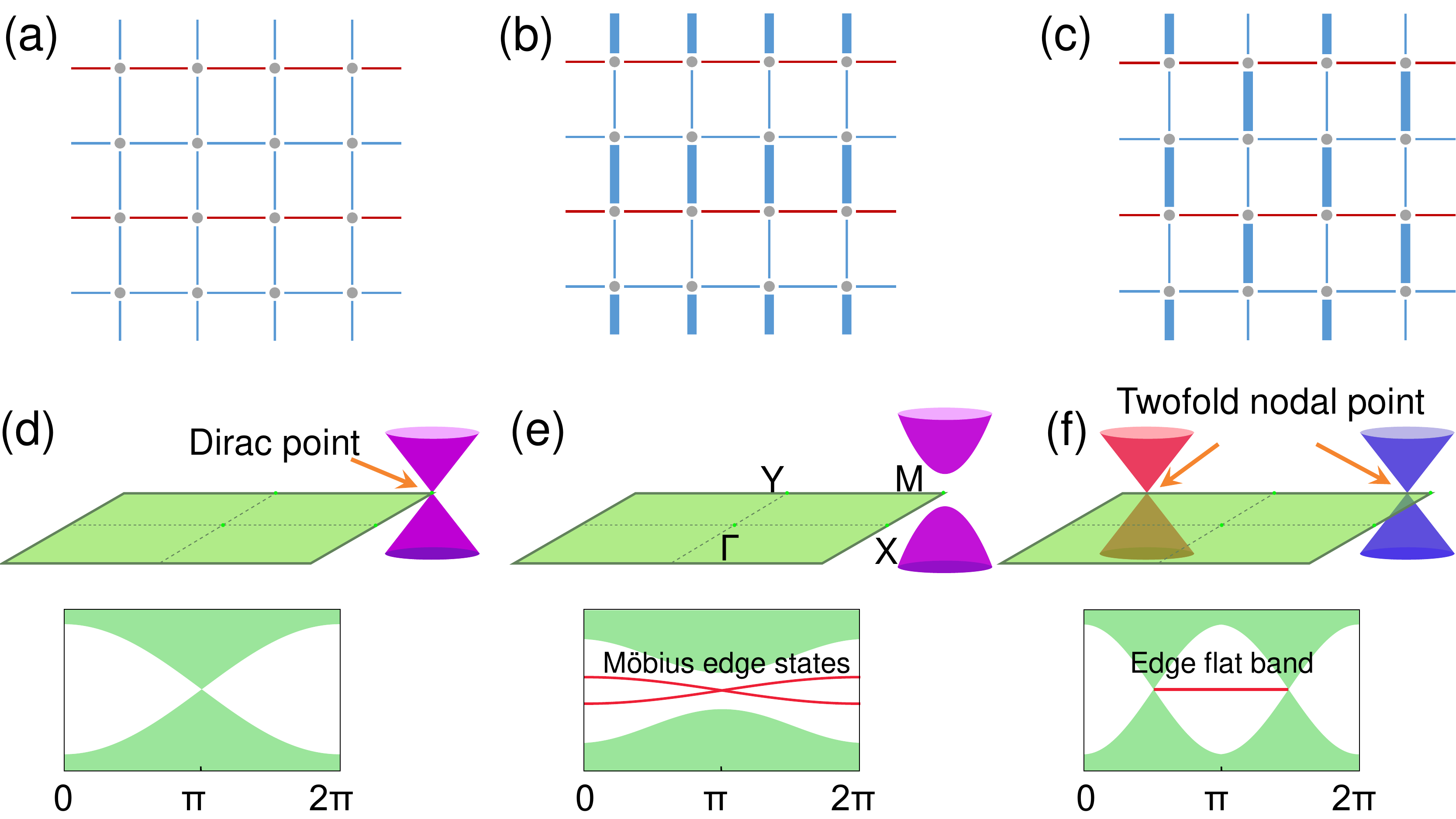}
  \caption{Illustration of the models with projective translation symmetries. The red (blue) bonds denote negative (positive) hopping amplitudes. Thus, each plaquette has a $\pi$ flux. (a) Both primitive translational symmetries $\L_{x,y}$ are preserved. The band structure features a fourfold degenerate Dirac point as shown in (d).  (b) The staggered dimerization pattern which preserves $\L_x$. The edge bands with the Mobius twist are illustrated in (e). (c) The alternative dimerization pattern with both primitive translation symmetries broken. It generates a graphene-like semimetal phase as illustrated in (f). The bottom panel of (d-f) illustrates the spectra for an open edge along $x$.}
  \label{fig1}
\end{figure*}

To demonstrate the idea, let us consider the simple 2D lattice model in Fig.~\ref{fig1}. Here, each rectangular plaquette carries a $\pi$ gauge flux. Fig.~\ref{fig1}(a) shows a specific gauge configuration, where each red (blue) colored bond has a negative (positive) hopping amplitude. The fundamental symmetries for the 2D lattice are the two primitive translations $\L_x$ and $\L_y$. Without the gauge field, the translations commute with each other, which is the algebra forming the foundation of solid state physics. However, one notes that under the $\Z_2$ gauge field, this fundamental algebra is modified to
\begin{equation}\label{Anti-commutation}
	\{\L_x,\L_y\}=0,
\end{equation}
because moving around a plaquette will endow the wave function with a $\pi$ phase. An immediate consequence is that each band is twofold degenerate for a generic momentum, since Eq.~\eqref{Anti-commutation} resembles the algebra of the Pauli matrices. More interestingly, at point M $(\pi,\pi)$ of the Brillouin zone (BZ), $\L_x$, $\L_y$, $T$ together with the imaginary unit $i$ generate a real Clifford algebra $C^{0,4}$, which has a unique four-dimensional representation (see Methods for more information). This means that the system must have a fourfold degenerate Dirac point at M (see Fig.~\ref{fig1}(d)), described by the Dirac model:
\begin{equation}\label{eq2}
	h_D(\bm{q})=q_x\Gamma_1+q_y\Gamma_2.
\end{equation}
Here, $\Gamma_\mu$ with $\mu=1,2,\cdots,5$ are the five Hermitian $4\times 4$ Dirac matrices satisfying $\{\Gamma_\mu,\Gamma_\nu\}=2\delta_{\mu\nu}1_4$. A concrete representation may be given by $\Gamma_1=\tau_3\otimes\sigma_2$, $\Gamma_2=\tau_2\otimes\sigma_0$, $\Gamma_3=\tau_1\otimes\sigma_0$, $\Gamma_4=\tau_3\otimes\sigma_1$ and $\Gamma_5=\tau_3\otimes\sigma_3$,  with $\tau$'s and $\sigma$'s being two sets of the Pauli matrices. It is important to note that this Dirac point is enabled solely by the projective translation symmetries and $T$, and there is only a single Dirac point in the BZ, which contrasts with all previous cases where Dirac points must require additional point group symmetries and they cannot exist as a single Fermi point in 2D $T$-invariant systems \cite{Young2015a}.

Breaking the primitive translation such as  $\L_y$ by dimerization will destroy the Dirac point and drive a topological phase transition. Two representative configurations are shown in Figs.~\ref{fig1}(b) and \ref{fig1}(c). In the Dirac model in Eq.~\eqref{eq2}, the two dimerization patterns correspond to perturbation terms $m_1\Gamma_3$ and $m_2 i\Gamma_2\Gamma_5$, respectively.

Interestingly, the case in Fig.~\ref{fig1}(b) realizes a M\"{o}bius topological insulator phase. In the bulk, a band gap is fully open and the band structure is characterized by a M\"{o}bius $\Z_2$ topological invariant enabled by $\L_x$ and the sublattice symmetry $\Gamma_5$. {In the eigenspace of $\L_x$, the Hamiltonian $H(\bm{k})$ is diagonalized into two blocks: $H(\bm{k})=\mathrm{diag}(h_1(\bm{k}),h_2(\bm{k}))$, which are connected by the sublattice symmetry $\Gamma_5$. The topological invariant is given by~\cite{zhao2020}
\begin{equation}
	\nu=\frac{1}{2\pi}\int_{[0,2\pi)\times S^1} d^2k~ \mathcal{F}+\frac{1}{\pi}\gamma(0)\mod 2.
\end{equation}
Here, $\mathcal{F}$ is the Berry curvature, and $\gamma(0)$ is the Berry phase on the $k_x=0$ path in the BZ. Both are defined for the valence bands of $h_1(\bm{k})$.} The hallmark of this insulator is that its edge parallel to the $x$ direction will have a M\"{o}bius edge band, i.e., the band has a twisted structure similar to the edge of a M\"{o}bius strip (see Fig.~\ref{fig1}(e)). Furthermore, the band is completely detached from the bulk bands, distinct from the usual topological insulators where the edge band must connect the bulk conduction and valence bands. We note that similar M\"{o}bius states were discussed in a few limited cases but all require complicated nonsymmorphic symmetries \cite{shiozaki2015, zhao2016, chang2017, zhang2020}, in contrast to the case here based on the projective translation symmetry. And the M\"{o}bius edge states have never been experimentally demonstrated before.

\begin{figure*}
  \centering
  \includegraphics[width=\textwidth]{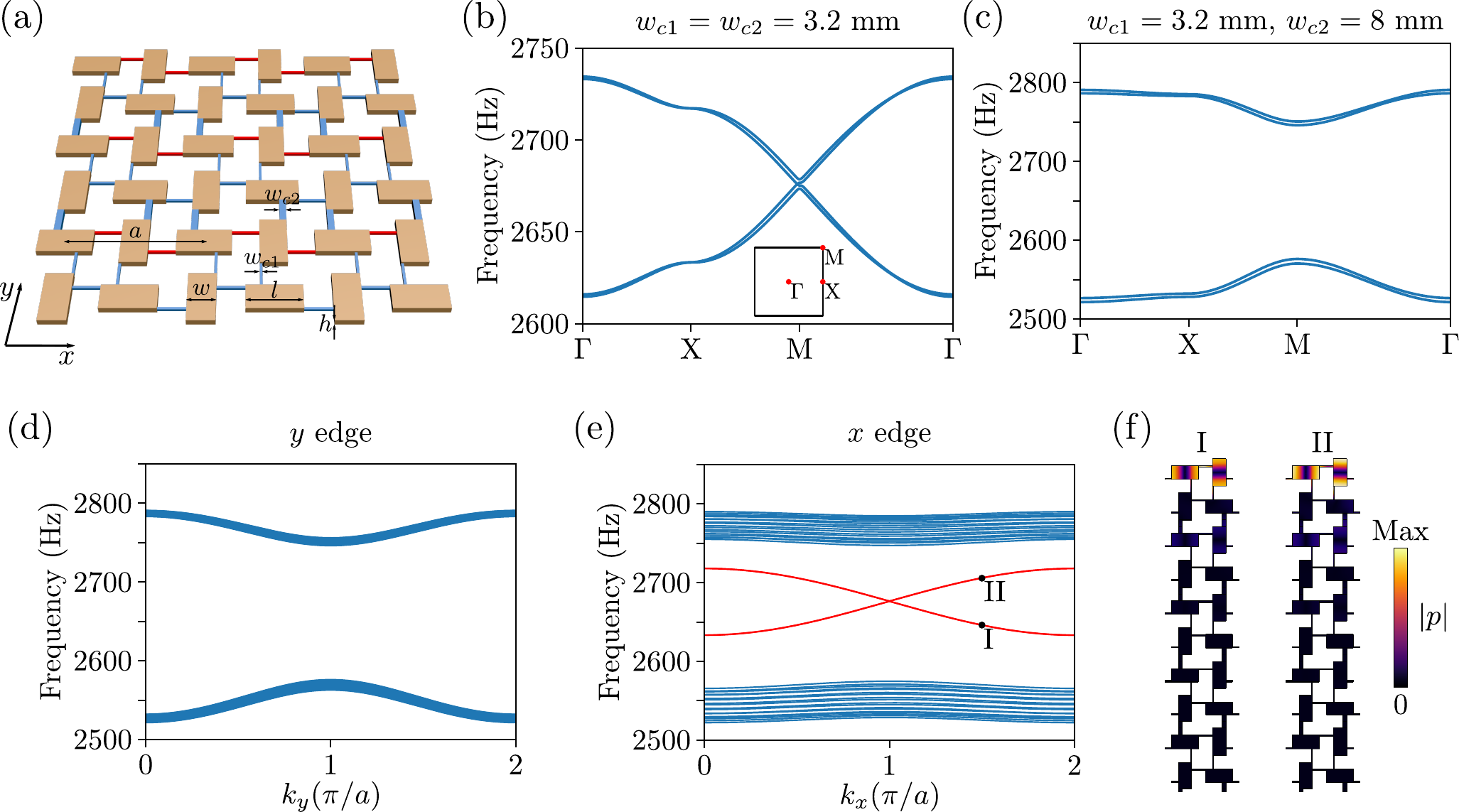}
  \caption{Simulations of an acoustic lattice hosting $\Z_2$-projective translational symmetry protected topological phase. (a) Schematic of designed acoustic lattice, which is made by cuboid resonators (orange) and coupling tubes (red and blue). The lattice constant $a=160$ mm. Other parameter values are given in the main text. (b)-(c) Simulated bulk bands for $w_{c1}=w_{c2}=3.2$ mm (b) and $w_{c1}=3.2$ mm, $w_{c2}=8$ mm (c), respectively. (d)-(e) Simulated dispersions for structures with the open boundary conditions along the $x$ direction but periodic boundary conditions along the $y$ direction, and vice versa, respectively. Bands colored in blue (red) correspond to bulk (edge) modes. In both simulations, the lattice consists of ten unit cells. (f) Eigenmode patterns for the two edge modes marked by ``I" and ``II" in (e). Note in (f) only part of the lattice is shown for clarity. There are negligible fields distributed in the area that is not shown.}
  \label{fig2}
\end{figure*}

The alternative dimerization in Fig.~\ref{fig1}(c) splits the fourfold Dirac point into two twofold nodal points along the $k_y$ direction. While both primitive translational symmetries $\L_x$ and $\L_y$ are broken, the sublattice symmetry $\Gamma_5$ is preserved. Hence, there is a topological charge defined by the winding number $w=\frac{1}{4\pi i}\oint_{C} d\bm k\cdot \mathrm{tr}\Gamma_5 H^{-1}(\bm{k})\nabla H(\bm{k})$ on a circle $C$ surrounding a Fermi point. The nontrivial topological charge leads to a flat edge band on the edge parallel to $x$, connecting the projected images of the two Fermi points (Fig.~\ref{fig1}(f)). One may note that this phase is similar to graphene, but there are actually important differences. First, twofold linear nodal points are known to be common to hexagonal lattices (including graphene), but here they occur in a rectangular lattice. It appears that the $\pi$ flux effectively transforms the topology of a rectangular lattice into that of a hexagonal lattice. Second, the points in graphene are pinned at the high-symmetry points $K$ and $K'$, whereas the points here are unpinned, i.e., they can freely move on the Y-M path without breaking any symmetry. Indeed, by tuning only the hopping strengths of  red-colored bonds, the two Fermi points can be continuously merged into the Dirac point and then annihilated, with the system transitioned into the M\"{o}bius insulator phase.

Without dimerization (Fig.~\ref{fig1}(a)), the primitive unit cell actually consists of two sites. Then, there are two twofold Dirac points at $(\pm \frac{\pi}{2},\pi)$, and the fourfold Dirac point for the doubled unit cell is actually folded from the two twofold Dirac points. We have chosen the doubled unit cell because we wanted to discuss the criticality of topological phases from two dimerization patterns. Another reason for adopting the doubled unit cell is that the BZ for the primitive unit cell has a $\pi$-periodicity for $k_x$, which can be understood as following. The BZ is specified by the commuting translation operators $L_y^2=e^{ik_y}$ and $L_x=e^{ik_x}$, and therefore $L_y$ is a nontrivial symmetry operator. Then, $L_y\psi(\bm{k})$ have the same energy of $\psi(\bm{k})$ but  $L_xL_y\psi(\bm{k})=e^{i(k_x+\pi)}L_y\psi(\bm{k})$.

One may wonder, as the primitive unit cell have already corresponded to a twofold Dirac semimetal phase, why we make the alternative dimerization. The answer is without dimerization the twofold Dirac points have no topological charge, and therefore there is no edge flat band. Thus, it is remarkable that the graphene-like topological semimetal can be realized on a rectangular lattice. Meanwhile, it is noteworthy that here the projectively represented translation symmetries play an essential role, but for graphene the twofold degeneracy is inherited from the $D_3$ symmetry.

\begin{figure*}
  \centering
  \includegraphics[width=\textwidth]{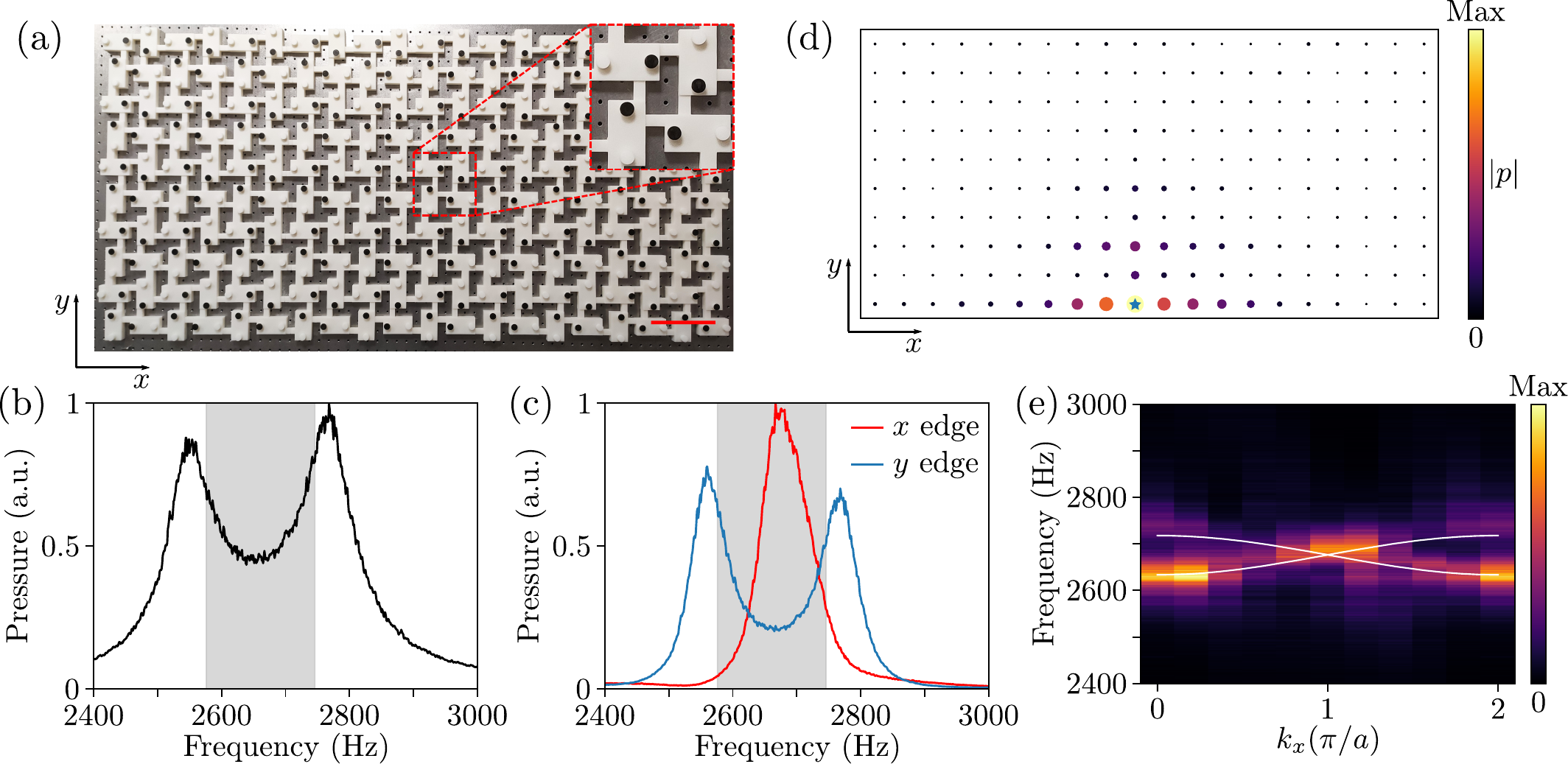}
  \caption{Experimental observation of M\"{o}bius edge modes. (a) A photo of fabricated sample with 20 $\times$ 10 resonators. The inset shows one unit cell. The scale bar is 160 mm. (b) Measured acoustic pressure when the speaker and microphone are placed at two different resonators inside the bulk. The grey region denotes simulated bulk bandgap. (c) Measured acoustic pressure when the speaker and microphone are placed at two different resonators on the $x$ edge (red curve) and the $y$ edge (blue curve). The grey region denotes simulated bulk bandgap. (d) Measured acoustic pressure distribution at 2680 Hz. Both color and size of the circles denote measured amplitude. (e) Measured edge dispersion from Fourier transformation of the measured fields on the bottom edge. Here the white lines indicate simulated edge bands.}
  \label{fig3}
\end{figure*}

Now we proceed to implement the above described projective translation symmetry protected topological phases in an acoustic crystal.  Our design, as depicted in Fig.~\ref{fig2}(a), consists of cuboid acoustic resonators (colored in orange) and coupling tubes (colored in red and blue). This kind of design has been used to construct various topological tight-binding Hamiltonians in acoustics, such as Weyl semimetals \cite{xiao2015, yang2016, li2018, ge2018, peri2019} and higher-order topological insulators \cite{xue2019a, ni2019, xue2019b, weiner2020, xue2020, ni2020, qi2020}. However, the underlying projective crystal symmetries and their resultant topological physics were never revealed. Here, the cuboid resonator has a size of 64 mm $\times$ 32 mm $\times$ 8 mm, supporting a dipolar mode at around 2680 Hz. Couplings between the resonators are enabled by thin tubes with square cross-sections. The whole structure is hollow, filled with air and surrounded by hard walls.

It is significant to note that the sign of coupling can be easily controlled by tuning the position of a coupling tube. Positive and negative couplings are realized by placing the coupling tubes at different sides of the dipolar mode's nodal line. In Fig.~\ref{fig2}(a), tubes that enable positive and negative couplings are colored in blue and red, respectively. In such configuration, this acoustic lattice carries $\pi$ flux per plaquette, as required. Moreover, the coupling strengths can be engineered by tuning the widths of the coupling tubes. Thus, the tight-binding model with both primitive translation symmetries $\L_{x,y}$, as well as two aforementioned dimerization patterns, can all be realized.

We performed a series of finite-element simulations with COMSOL Multiphysics to investigate the physics of this acoustic lattice. In the absence of coupling dimerization (i.e., all the coupling tubes have the same width), the bulk bands are all nearly two-fold degenerate over the entire BZ except for M point, where a fourfold Dirac point approximately appear, as can be seen from Fig.~\ref{fig2}(b). These features are consequences of the projectively represented translation symmetries and $T$, as discussed above.

Then, we proceed to transition the fourfold degenerate Dirac criticality into other topological phases by introducing dimerization. We impose a staggered dimerization pattern (see Fig.~\ref{fig2}(a)) by letting $w_{c1}=3.2$ mm and $w_{c2}=8$ mm. Since the primitive translation symmetry $\L_y$ is broken, the degeneracy at M is lifted and a band gap is opened, as shown in Fig.~\ref{fig2}(c). Next, we look into the boundary modes in this gapped phase. On the open edge parallel to $x$, one clearly observes two crossing edge bands inside the bulk bandgap, and particularly fully detached from the bulk bands (see Fig.~\ref{fig2}(e)). These edge bands, as discussed previously, actually represent a single band forming a M\"{o}bius twist. In contrast, for the open edge along the $y$ direction, $\L_x$ is broken and therefore  no in-gap edge modes are observed (see Fig.~\ref{fig2}(d)). All these simulation results agree well with the predicted phenomena (see Supplementary Information for calculations on tight-binding models).

Having numerically confirmed the validity of our design, we then conduct experiments to probe the signatures of the M\"{o}bius topological insulator. As shown in Fig.~\ref{fig3}(a), a sample with 20 $\times$ 10 resonators was fabricated through 3D printing. On the top of each resonator, there are two small holes where excitation and detection can be conducted. When not in use, these holes are covered by plugs to prevent sound leakage (see the inset in Fig.~\ref{fig3}(a) for details; here the plugs in white (black) are used to cover the holes for excitation (detection)). We first measure the bulk transmission to confirm the existence of the bandgap. The measured acoustic pressure, as plotted in Fig.~\ref{fig3}(b), shows two peaks separated by a gap. The frequency range of measured gap matches well with the bandgap found in simulation (shaded in grey in Fig.~\ref{fig3}(b)). Then, we measure the transmission on the edges of the sample. For the $y$ edge, the measured spectrum (blue curve in Fig.~\ref{fig3}(c)) is similar to the bulk transmission spectrum. We only observed two separated peaks lying at frequencies corresponding to bulk bands, which is consistent with the simulation given in Fig.~\ref{fig2}(d) showing that there are no in-gap edge modes. In contrast, the transmission spectrum obtained on $x$ edge is significantly different from the previous two. As given by the red curve in Fig.~\ref{fig3}(c), there is only one peak in the spectrum. Furthermore, this peak is inside the bandgap, consistent with the consequence of the M\"{o}bius-twisted edge band on the $x$ edge.

In addition to the transmission measurements as described above, we also mapped out the field distribution to further confirm the existence of the M\"{o}bius edge band. In this experiment, a speaker was fixed at the center of the bottom edge (denoted by the blue star in Fig.~\ref{fig3}(d)) and a microphone scanned the whole sample to obtain both amplitude and phase of sound at each resonator (See Supplementary Information for details). Fig.~\ref{fig3}(d) shows the measured pressure at 2680 Hz (corresponding to the peak frequency of the $x$-edge spectrum in Fig.~\ref{fig3}(c)). As can be seen, the sound wave is mainly localized at the edge. The pressure at both sides of the source is found to be similar, which reflects the fact that edge bands have both positive group velocity and negative velocity branches. Besides, we can see from Fig.~3(d) that the measured pressure is higher in the third row than in the second row (counting from the bottom edge), because the first and third rows belong to the same sublattice which is excited more than the other sublattice. This is consistent with the eigenmode pattern found in the simulation (see Fig.~\ref{fig2}(f)). We note that the excited edge modes decay fast along the edge due to the background loss caused by material absorption, which is a common issue in coupled acoustic resonator lattices. Furthermore, the edge dispersion can be directly visualized by performing Fourier transformation on the measured field distribution data. As can be seen from Fig.~\ref{fig3}(e), the Fourier spectrum (colormap) agrees well with the simulated edge dispersion (solid white lines). These experiment results in both real space and momentum space, together with the transmission spectra, confirm the existence of the M\"{o}bius edge band.

\begin{figure}
  \centering
  \includegraphics[width=\columnwidth]{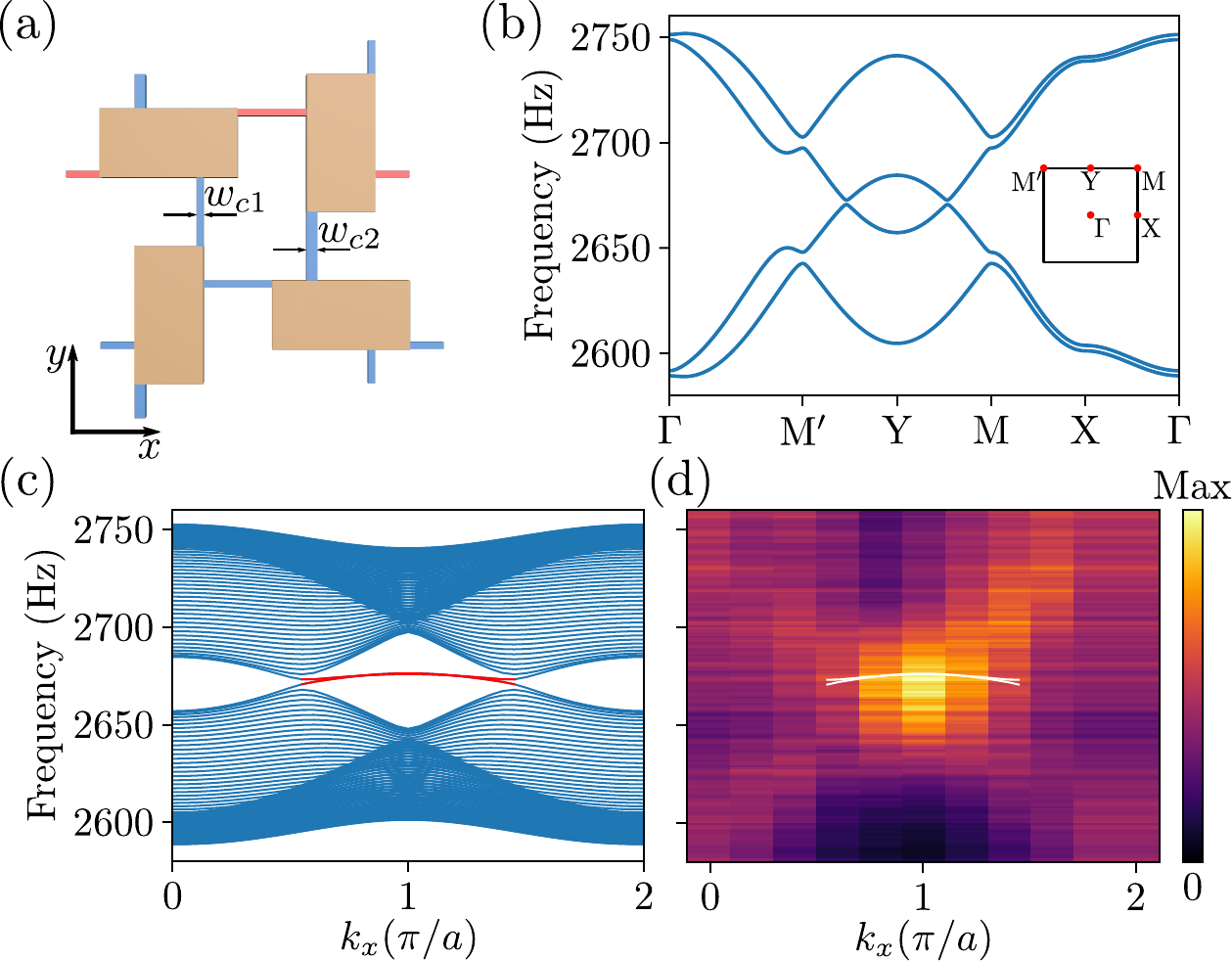}
  \caption{Experimental demonstration of a graphene-like semimetal. (a) Unit cell of the acoustic lattice with alternative dimerization. (b) Simulated bulk bands for $w_{c1}=3.2$ mm and $w_{c2}=5$ mm. (c) Simulated dispersions for a lattice with the open boundary conditions along the $y$ direction but periodic boundary conditions along $x$ direction. There are edge states (colored in red) connecting the projections of the Fermi points. (d) Experimentally measured edge dispersion. Here the white lines denote simulated edge bands.}
  \label{fig4}
\end{figure}

In above simulations and experiments, we have shown that a staggered dimerization can transform the acoustic Dirac semimetal into a  M\"{o}bius topological insulator. We also implement a different dimerization pattern, following that in Fig.~\ref{fig1}(c), which leads to a graphene-like semimetal phase. The acoustic unit cell with this alternative dimerization is shown in Fig.~\ref{fig4}(a), where the coupling tubes enabling weak and strong couplings have widths $w_{c1}=3.2$ mm and $w_{c2}=5$ mm, respectively. The corresponding bulk dispersion, as given in Fig.~\ref{fig4}(b), clearly shows the emergence of two two-fold nodal points on M$'$-M. As discussed above, these two Fermi points, similar to the ones in graphene, carry nontrivial topological charges that lead to edge bands connecting the projected images of the two Fermi points. To see this, we simulated a strip with periodic boundary conditions along $x$ and open boundary conditions  along $y$ (40 cells). As shown in Fig.~\ref{fig4}(c), we indeed observe edge bands (red curves) connecting the projections of the Fermi points. We again conducted acoustic pressure measurement on the edge resonators under an edge excitation. The resulted Fourier spectrum is plotted in Fig.~\ref{fig4}(d). As can be seen, the frequencies and momenta of high intensity regions match well with the edge band in Fig.~\ref{fig4}(c) (which is also plotted in Fig.~\ref{fig4}(d) as the white curve).

In conclusion, we demonstrated a new class of topological phases protected by projective crystal symmetries in an acoustic system. A Dirac semimetal enforced by projective translation symmetries was shown to give rise to a M\"{o}bius insulator and a graphene-like semimetal under different perturbations. Our results point to a promising yet unexplored direction to discover novel topological phases. There are various ways to generalize the results in this work. Firstly, while our demonstration is in acoustics, the idea can also be realized in other classical systems like photonic/phononic crystals \cite{keil2016, serra2018, peterson2018} and electric circuits \cite{imhof2018}. Secondly, it would be interesting to explore more possibilities by increasing system dimension and considering other types of projective symmetries \cite{zhao2021, shao2021}. Besides, our system also provides a natural platform to study the effect of non-Hermiticity through introducing dissipation \cite{gao2021}.

\bigskip
\noindent{\large{\bf{Methods}}}

\noindent\textbf{The Clifford algebra for the fourfold degenerate Dirac point.} Since time-reversal $T$ is anti-unitary, we should treat the imaginary unit $i$ as an operator. Then, we recombine the operators $\L_{x,y}$, $T$ and $i$ into the following four operators:
\begin{equation}
i\L_x,\quad  i\L_y,\quad T, \quad iT,
\end{equation}
which anti-commute with each other. At M=$(\pi,\pi)$, we have $(\L_{x,y})^2=-1$. Then, the squares of the operators are given by
\begin{equation}
(i\L_x)^2=1, \quad (i\L_y)=1, \quad T^2=1, \quad (iT)^2=1.
\end{equation}
Thus, the four operators generate a Clifford algebra with all generators squared to 1. The Clifford algebra is denoted by $C^{0,4}$. This Clifford algebra is known to have a unique four-dimensional representation~\cite{Clifford}.

\noindent\textbf{Numerical and experimental details.} Numerical simulations presented in this work are performed by COMSOL Multiphysics, pressure acoustics module. The interfaces between air and the 3D printing materials (photosensitive resin) are modelled as sound rigid walls due to the large impedance mismatch. In all simulations, we have set sound speed and density of air to be 343 m/s and 1.3 kg/m$^3$, respectively.

The samples used in experiments are fabricated through a stereolithography apparatus with a resolution around 0.1 mm. The thickness of the photosensitive resin is 5 mm. There are two circular holes (radius = 5 mm) on each resonator for excitation and detection. When not in use, they are covered by plugs.

In experiments, the acoustic signal is generated by a speaker (Tymphany PMT-40N25AL01-04) and detected by a microphone (Br\"{u}el \& Kj{\ae}r Type 4182). The measured signal is processed by an analyser (Br\"{u}el \& Kj{\ae}r 3160-A-022 module) to get the frequency spectrum. See Supplementary Information for more details on the experiment setup.

\bigskip
\noindent{\large{\bf{Data availability}}}

\noindent The experimental data are available in the data repository for Nanyang Technological University at this link (URL to be inserted upon publication). Other data that support the findings of this study are available from the corresponding authors on reasonable request.

\bigskip
\noindent{\large{\bf{Acknowledgements}}}

\noindent We are grateful to Hongxiang Sun and He Gao for helpful discussions on experimental measurements. This work is supported by the Singapore Ministry of Education Academic Research Fund Tier 3 Grant MOE2016-T3-1-006, and Tier 2 Grant MOE2019-T2-2-085. 

\bigskip
\noindent{\large{\bf{Author contributions}}}

\noindent Y.Z., S.Y. and B.Z. conceived the idea and supervised the project. Y.Z., Y.H. and S.Y. did the theoretical analysis. H.X. performed the simulations and designed the sample. H.X., Z.W., Z.C., L.Y. and Y.F. conducted the experiments. H.X., Y.Z., S.Y. and B.Z. wrote the manuscript with input from all authors.

\bigskip
\noindent{\large{\bf{Competing interests}}}

\noindent The authors declare no competing interests.
\end{document}